\documentclass[aps,prl,preprint,showpacs,groupedaddress]{revtex4}
\usepackage{amsmath}
\usepackage{tipa}
\usepackage{bbm}
\usepackage{txfonts}
\usepackage{graphicx}  
\usepackage{dcolumn}   
\usepackage{bm}        
\usepackage{amssymb}   
\usepackage{latexsym}   

\begin{document}

\preprint{Submitted to PRL}

\title{Test of the universality of free fall with atoms in different spin Orientations}

\author{Xiao-Chun Duan}
\author{Min-Kang Zhou}
\author{Xiao-Bing Deng}
\author{Hui-Bin Yao}
\author{Cheng-Gang Shao}
\author{Jun Luo}
\author{Zhong-Kun Hu}\email[E-mail: ]{zkhu@mail.hust.edu.cn}

\affiliation{MOE Key Laboratory of Fundamental Physical Quantities
Measurements, School of Physics, Huazhong University of Science and
Technology, Wuhan 430074, People's Republic of China}

\date{\today}

\begin{abstract}

We report a test of the universality of free fall (UFF) related to
spin-gravity coupling effects by comparing the gravity acceleration
of the $^{87}$Rb atoms in $m_F=+1$ versus that in $m_F=-1$, where
the corresponding spin orientations are opposite. A
Mach-Zehnder-type atom interferometer is exploited to sequentially
measure the free fall acceleration of the atoms in these two
sublevels, and the resultant E$\rm{\ddot{o}}$tv$\rm{\ddot{o}}$s
ratio determined by this work is ${\eta _S} =(-0.2\pm1.5)\times
10^{-5}$. The interferometer using atoms in $m_F=+1$ or $m_F=-1$ is
highly sensitive to magnetic field inhomogeneity, which limits the
current experimental precision of our UFF test. The work here
provides a stepping stone for future higher precision UFF test
related to different spin orientations on atomic basis.

\end{abstract}

\pacs{37.25.+k, 03.75.Dg, 04.80.Cc}

\maketitle

The universality of free fall (UFF) is one of the fundamental
hypotheses in the foundation of Einstein's general relativity (GR)
\cite{1}, which states that all test bodies fall with the same
acceleration in the gravitational field regardless of their
structure and composition. Traditional verifications of the UFF are
performed with macroscopic bodies that weight differently or
comprise of different material by torsion balance technique
\cite{2,3,4}, free-fall method \cite{5,6,7} or laser ranging mission
\cite{8,9}, achieved a level of 10${}^{-13}$ \cite{3,4,9}. There are
also lots of work investigating possible violation of UFF that may
be induced by spin-related forces (see, for example,
\cite{10,11,12,13,14,15}), and UFF tests of this kind have been
performed with polarized or rotating macroscopic bodies
\cite{16,17,18,19,20,21,22,23,24}. Here we report a spin-orientation
related UFF test with quantum objects by atom interferometry.

UFF tests with quantum objects have earlier been performed with a
neutron interferometer \cite{25}, and in recent years, were carried
out by comparing the free fall acceleration between different atoms
or between atoms and macroscopic masses \cite{26,27,28,29,30,31}.
The motivation of using quantum objects is not only for potentially
higher precision or associated well defined properties, but also for
more possibilities to break Einstein equivalence principle on
quantum basis \cite{32}. For example, the variation of the free fall
acceleration with atoms in different hyperfine levels has also been
tested in Ref. \cite{27} at a level of 10${}^{-7}$. Recently,
Tarallo et al.  \cite{33} performed an UFF test using the bosonic
$^{88}$Sr isotope ($I=0$) and the fermionic $^{87}$Sr isotope
($I=9/2$) at a level of 10$^{-7}$ by Bloch oscillation. In their
experiment, the $^{87}$Sr atoms were in a mixture of different
magnetic sublevels, resulting in effective sublevel of $\left\langle
{{m_F}} \right\rangle = 0$. They also gave an upper limit on the
spin-gravity coupling by analyzing the broadening caused by possible
different free fall accelerations between different magnetic
sublevels. However, we note that possible anomalous spin-spin
couplings (see, for example, \cite{20,34,35}) or dipole-dipole
interaction (see, for example, \cite{36}) between the $^{87}$Sr
atoms with different magnetic sublevels may disturb, or even cover
the spin-gravity coupling effects in their experiment. Since most
models describing spin-gravity coupling imply a dependence on the
orientation of the spin, we perform a new UFF test with ${}^{87}$Rb
atoms sequentially prepared in two opposite spin orientations (Fig.
1), namely $m_F=+1$ versus $m_F=-1$. The corresponding free fall
accelerations are compared by atom inteferometry
\cite{37,38,39,40,41}, which determines the spin-orientation related
E$\rm{\ddot{o}}$tv$\rm{\ddot{o}}$s ratio \cite{42} as
\begin{equation}
{\eta _S} \equiv 2\frac{{{g_{ + }} - {g_{ - }}}}{{{g_{ + }} + {g_{ -
}}}}. \label{Eq:1}
\end{equation}
In Eq.(1), the gravity acceleration of atoms in $m_F=+1$ ($m_F=-1$)
is denoted as $g_{+}$ ($g_{-}$) to account for possible difference.
This provides a direct way to test spin-orientation related UFF on
quantum basis.

\begin{figure}[tbp]
\includegraphics[trim=40 10 40 20,width=0.40\textwidth]{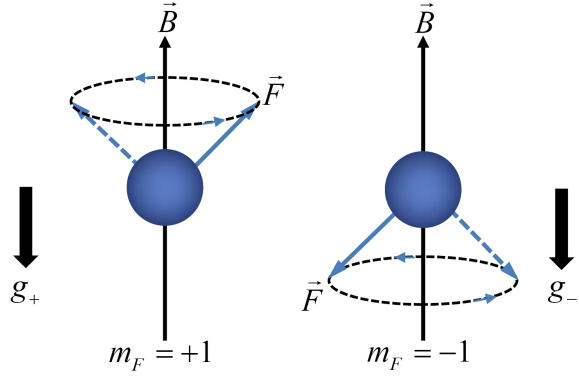}
\caption{\label{fig:1}(color online) Schematic of the spin
orientations for ${}^{87}$Rb atoms in magnetic sublevels $m_F=+1$
versus $m_F=-1$ of the 5${}^{2}$S${}_{1/2}$ hyperfine levels. The
bias magnetic field $\vec B$ defines the external direction to which
the atoms spin is referenced. And the total angular momentum of each
atom (denoted by the $\vec F$) processes around $\vec B$.}
\end{figure}

Compared with UFF tests using polarized or rotating macroscopic
masses, it is much simpler to prepare cold atomic ensemble with pure
polarization using stimulated Raman transition \cite{43}. However,
with atoms in sublevels $m_F=+1$ or $m_F=-1$, the Zeeman effect is
considerable, which makes the interferometer highly sensitive to the
magnetic field inhomogeneity. What's worse is that the phase shift
induced by the inhomogeneity is opposite for atoms in the two
sublevels, which thus can't be directly canceled in the final
comparison of measured free fall accelerations. For the
interferometer with atoms in $m_F$, the phase shift induced by the
gravitational acceleration and the magnetic field gradient is
expressed as
\begin{equation}
\varphi _{{m_F}}^ \pm  =  \mp
{k_{{\rm{eff}}}}{g_{{m_F}}}T_{{\rm{eff}}}^2 + 2{\alpha
_{Z,I}}{m_F}{\gamma _B}({V_\pi } \mp {V_r}/2){T^2},\label{Eq:2}
\end{equation}
where the superscript $\pm$  denotes the corresponding direction of
$\emph{\textbf{k}}_{\rm{eff}}$ in the interfering process, with
$+k_{{\rm{eff}}}$ ($-k_{{\rm{eff}}}$) indicating the same (opposite)
directions between $\emph{\textbf{k}}_{\rm{eff}}$ and local
gravitational acceleration. And $T$ is the separation time between
Raman laser pulses, while ${T_{{\rm{eff}}}} \equiv T\sqrt {1 + 2\tau
/T + 4\tau /\pi T + 8{\tau ^2}/\pi {T^2}}$ is the effective
separation time accounting for the effect of finite Raman pulses
duration ( $\tau$ is the duration for the $\pi/2$ Raman pulse)
\cite{37}. In Eq.(2), the second term corresponds to that induced by
the magnetic field (only the first order of the inhomogeneity is
considered), where $\gamma_B$ is the magnetic field gradient,
$\alpha_{Z,I}$ is the strength of first-order Zeeman shift for
$^{87}$Rb atoms in 5$^2S_{1/2}$ state, $V_r$ is the recoil velocity,
and $V_{\pi}$ is the average vertical velocity of the atoms in $F=1$
at the moment of the interfering $\pi$ pulse (in this work, the
atoms are initially prepared in $F=1$ before the interfering).

In order to alleviate the influence of the magnetic field
inhomogeneity, according to Eq. (2), three steps are taken in this
work. Firstly, the magnetic field throughout the interfering space
is mapped \cite{44,45}, and a region where the field is relative
homogeneous is selected for the interfering to take place, namely
making $\gamma_B$ as small as possible. The selected region is at
about 736 mm height above the magnetic-optical trap (MOT) center.
And there the magnetic field varies less than 0.1 mG over
several-millimeters vertical distance, while the magnitude of the
bias magnetic field is about 115 mG. Secondly, the direction of the
effective Raman laser wave number $\emph{\textbf{k}}_{\rm{eff}}$ can
be reversed to make a differential measurement for each $m_F$
\cite{37}. A majority of the influence (the part associated with
$V_{\pi}$ in Eq. (2)) induced by the magnetic gradient will be
canceled using this differential measurement, since the influence is
almost independent off $\emph{\textbf{k}}_{\rm{eff}}$ (we note that
$V_{\pi}$ is typically much larger than $V_r$). However, with the
Raman lasers configured in $+k_{{\rm{eff}}}$ versus
$-k_{{\rm{eff}}}$, the directions of the recoil velocities are
opposite. This induces a tiny difference between the atoms'
trajectories. And consequently causes a residual influence in the
differential measurement result, which is the part associated with
$V_r$ in Eq. (2). The third step is to correct this residuum using
the $\gamma_B$ obtained from the common mode result for the two
interfering configurations of $\emph{\textbf{k}}_{\rm{eff}}$.
According to Eq.(2), for each $m_F$, the differential mode
measurement result ($\Delta \varphi _{{m_F}}^d \equiv (\Delta
\varphi _{{m_F}}^ +  - \Delta \varphi _{{m_F}}^ - )/2$) and the
common mode measurement result ($\Delta \varphi _{{m_F}}^c \equiv
(\Delta \varphi _{{m_F}}^ +  + \Delta \varphi _{{m_F}}^ - )/2$) are
respectively
\begin{equation}
\left\{ \begin{array}{l} \Delta \varphi _{{m_F}}^d = -
{k_{{\rm{eff}}}}{g_{{m_F}}}T_{{\rm{eff}}}^2 - {\alpha _{Z,I}}{m_F}{\gamma _B}{V_r}{T^2}\\
\Delta \varphi _{{m_F}}^c = 2{\alpha _{Z,I}}{m_F}{\gamma _B}{V_\pi
}{T^2}
\end{array} \right..\label{Eq:3}
\end{equation}
According to Eq.(3), the magnetic field gradient $\gamma_B$ can be
directly estimated from $\Delta \varphi _{{m_F}}^c$, as long as
$V_{\pi}$ is known.

In order to maximally suppress the magnetic-field-gradient influence
in the differential measurement for each $m_F$, it is required that
both $\gamma_B$ and $V_{\pi}$ are the same between the
$+k_{{\rm{eff}}}$ and $-k_{{\rm{eff}}}$ interfering configurations.
The two requirements can be simultaneously satisfied by preparing
the atomic ensembles in the same average velocity, namely $V_s^+  =
V_s^- $ ($V_s$ denotes the average velocity of the atomic ensemble
after the state preparation, and the superscript $\pm$ denotes the
$\emph{\textbf{k}}_{\rm{eff}}$ configuration). In this case, the
atomic ensembles are in the same region when the interfering is
taking place, and thus $\gamma_B$ is the same, while $V_{\pi}$ is
obviously the same since it is directly determined by $V_s$. Though
techniques are mature for preparing the atoms state by
velocity-sensitive Raman transition (VSRT) \cite{43}, it is in fact
not simple to ensure the equality between $V_s^+$ and $V_s^- $.
Using conventional state preparation method (see, for example,
\cite{46}), the equality will strongly depend on the pre-determined
Zeeman shift and AC-Stark shift. And the corresponding variations
will cause opposite changes for $V_s^+$ and $V_s^- $. Here we
explore an easy but reliable method to guarantee this equality. For
the two interfering configurations, we implement the state
preparations using the Raman lasers both configured in
$+k_{\rm{eff}}$ with the same effective frequency $\omega
_{{\rm{eff}}}$ (${\omega _{\rm{eff}}} \equiv {\omega _1} - {\omega
_2}$, namely the frequency difference of the two laser beams in
Raman lasers.). In this case, for each $m_F$, the state preparations
are completely the same for the two interfering configurations, and
thus the average velocities  of the selected atoms $V_s$ are
naturally the same. Compared with conventional operation of the
interferometer, in addition to usual Raman lasers frequency chirp,
this method needs an extra shift of $\omega _{{\rm{eff}}}$ after the
state preparation. This shift will switch the Raman lasers
configuration from $+k_{\rm{eff}}$ to $-k_{\rm{eff}}$ for the
interfering process where the Raman lasers need to be configured in
$-k_{\rm{eff}}$. This can be realized by using two arbitrary
function generators (AFG) to mix with a microwave signal source in
the Raman lasers' optical phase locking loop (OPLL), with one AFG to
implement the $\omega _{{\rm{eff}}}$ shift and the other to
implement the chirp.

The experiment is performed in an atom gravimeter that has been
previously reported in detail in Ref. \cite{40}. It takes 727 ms to
load about 10$^8$ cold $^{87}$Rb atoms from a dispenser using a
typical MOT. Then the atoms are launched upward and further cooled
to about 7 $\mu$K with a moving molasses procedure in the atomic
fountain. The apex of the fountain is at 750 mm height above the
MOT, close to the aimed interfering region at about 736 mm, which is
helpful to limit the atoms' flying distance during the interfering
process. After a flight time of 324 ms from the launch, a Raman
$\pi$ pulse with a duration of 26 $\mu$s is switched on to implement
the state preparation. With the Raman lasers configured in
$+k_{\rm{eff}}$, the detune (defined here as the difference between
$\omega _{{\rm{eff}}}$ and the the hyperfine splitting of the two
ground levels (5$^2S_{1/2}$) of the $^{87}$Rb atom) that selects the
maximum $m_F=+1$ atoms is found to be $-1546$ kHz, and that for
$m_F=-1$ is $-1866$ kHz. After the unwanted atoms are removed by a
blow-away beam, the atomic cloud arrives at 736mm height and
undergoes the $\pi/2-\pi-\pi/2$ Raman pulses with a pulse separation
time of $T=2$ ms. With larger $T$, the interferometry fringe would
become invisible due to different magnetic field inhomogeneity
experienced by respective atom in the ensemble. This is the reason
why the effect of the finite Raman pulses durations must be
considered in Eq. (2). The transition probability of the atoms after
the interfering is obtained through a normalized fluorescence
detection when the clouds falls back into the detection chamber. The
entire process of a single shot measurement as described above takes
1.5 s. Before the formal data acquisition, $V_{\pi}$  should be
measured to deduce $\gamma_B$ from $\Delta \varphi _{{m_F}}^c$. This
velocity can be obtained from the spectroscopy of the VSRT \cite{43}
with a Raman $\pi$ pulse applied at the right moment. Here the
spectroscopies with the Raman lasers configured in $+k_{\rm{eff}}$
and $-k_{\rm{eff}}$ are combined to make a differential measurement,
in which method the knowledge of the Zeeman shift or the AC-Stark
shift is not needed. The measured average velocity is
$V_{\pi}=509.0(1)$ mm/s for the selected atoms in $\left| {F =
1,{m_F} = + 1} \right\rangle$, and is $V_{\pi}=509.4(1)$ mm/s for
$\left| {F = 1,{m_F} = - 1} \right\rangle$ at the moment of the
interfering $\pi$ pulse.

\begin{figure}[tbp]
\includegraphics[trim=40 10 40 20,width=0.40\textwidth]{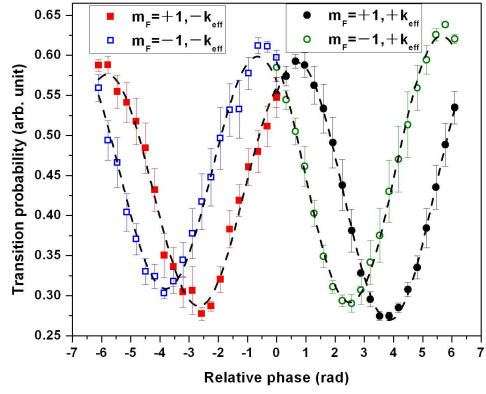}
\caption{\label{fig:2}(color online) Fringes for different
combinations of $m_F$ and $\emph{\textbf{k}}_{\rm{eff}}$, where each
fringe shown is an average of 10 fringes with one corresponding
combination. In one cycle, the fringes are obtained in turn for the
combinations of $m_F=+1$ and $+k_{\rm{eff}}$ (black circle),
$m_F=+1$ and $-k_{\rm{eff}}$ (red square), $m_F=-1$ and
$+k_{\rm{eff}}$ (olive empty circle), $m_F=-1$ and $-k_{\rm{eff}}$
(blue empty square).}
\end{figure}

Finally the measurement of the gravity acceleration of the atoms in
different magnetic sublevels is performed sequentially, with
different interfering configurations of the Raman lasers. One full
interferometry fringe is obtained by scanning the chirp rate of
$\omega _{{\rm{eff}}}$  in 20 steps for each $m_F$ in each
interfering $\emph{\textbf{k}}_{\rm{eff}}$ configuration, namely 30
s for a full fringe. Meanwhile, in order to reduce the effect of
possible drift of related quantities, for example, the Raman lasers
power, four adjacent fringes are grouped as a cycle unit, with one
fringe corresponding to one combination of $m_F$ and
$\emph{\textbf{k}}_{\rm{eff}}$. The switches between the
combinations are automatically controlled by the computer through
tuning the Raman lasers detune, and the typical fringes for the four
combinations are shown in Fig.2. The measurement is repeated about
28 hours from cycle to cycle, and the phase shifts are extracted by
the cosine fitting from the fringes. The differential mode result
and the common mode result are obtained from the combinations of the
corresponding phase shifts. The Allan deviation for the gravity
acceleration measurement is calculated from the differential mode
result for each $m_F$, which is shown in Fig.3, and the statistics
of the combined measurement results for each $m_F$ is also acquired
as shown in Table I. The Allan deviation for the measurement using
the atoms in $m_F=0$ with the Raman lasers configured in
$+k_{\rm{eff}}$ for $T=2$ ms is also shown in Fig.3 as a reference.
According to the Allan deviations, the short term sensitivity for
the interferometers using atoms in $m_F=\pm {1}$ is about 3.4$\times
10^{-3} g$/$\sqrt{\rm{Hz}}$, which implies a sensitivity of
2.4$\times 10^{-3}g$/$\sqrt{\rm{Hz}}$ if only one combination of
$m_F$ and $\emph{\textbf{k}}_{\rm{eff}}$ is consecutively repeated.
This induced sensitivity is about four times worse than that using
atoms in $m_F=0$, which is most probably due to the fluctuation of
the location where the atoms interacts with the interfering pulses
(and thus the fluctuation of the experienced magnetic field gradient
by the atoms). This fluctuation is caused by the variation of the
launch velocity as well as the initial launch position of the atomic
cloud from shot to shot.

\begin{figure}[tbp]
\includegraphics[trim=40 10 40 20,width=0.40\textwidth]{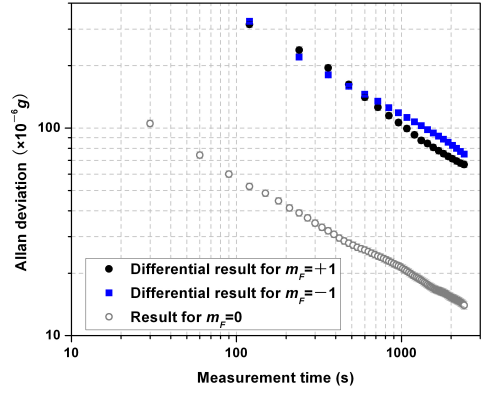}
\caption{\label{fig:3}(color online) Short-term Allan deviations for
the gravity acceleration measurements using atoms in different
$m_F$. The Allan deviations for $m_F=+1$ (black circle) and $m_F=-1$
(blue square) are calculated from the differential measurement
results, while that for $m_F=0$ (gray empty circle) is calculated
from the phase shifts obtained consecutively with Raman lasers
always configured in $+k_{\rm{eff}}$.}
\end{figure}

\newcommand{\tabincell}[2]{\begin{tabular}{@{}#1@{}}#2\end{tabular}}
\begin{table}[b]
\caption{\label{tab:table1} Statistics of the differential mode
measurement and the common mode measurement. The 2$\pi$ ambiguity is
easily removed in this work thanks to the rather short separation
time $T$. The magnetic field gradient is deduced from $\Delta
\varphi _{{m_F}}^c$, and the corrected $\Delta \varphi _{{m_F}}^{d}$
in the last column is the result of subtracting the residual
magnetic field inhomogeneity effect from the original $\Delta
\varphi _{{m_F}}^{d}$ in the second column.}
\begin{ruledtabular}
\begin{tabular}{ccccc}
\textrm{$m_F$}& \textrm{\tabincell{c}{$\Delta \varphi
_{{m_F}}^{d}$\\rad}}& \textrm{\tabincell{c}{$\Delta \varphi
_{{m_F}}^{c}$\\rad}}& \textrm{\tabincell{c}{$\gamma_B$\\$\mu$G/mm}}&
\textrm{\tabincell{c}{Corrected $\Delta \varphi
_{{m_F}}^{d}$\\rad}}\\
\colrule
$+$1 & $-$644.322(7) & $-$0.674(8) & $-$37.6(4)&$-$644.330(7)\\
$-$1 & $-$644.339(7) & $+$0.666(8) & $-$37.2(4)&$-$644.331(7)\\
\end{tabular}
\end{ruledtabular}
\end{table}

The uncertainties in Table I are the corresponding statistical
standard deviations. From the common mode results, the magnetic
field gradients experienced for atoms in each $m_F$ are deduced,
which are nearly equal as expected since the interfering region is
the same. According to Eq.(3), this gradient with a magnitude of
$-37 \mu$G/mm corresponds to a $+8$ mrad residual effect for the
differential result with atoms in $m_F=+1$, and $-8$ mrad residual
effect for $m_F=-1$. It shows that a majority of the phase shift due
to the magnetic field inhomogeneity is canceled in the differential
measurement, and the residual is only about $1.2\%$. And the
residual effect due to the Raman pulses durations is far less than
that level, which is thus safely neglected in this work. In this
differential measurement of the gravity acceleration with a rather
short separation time, some disturbances, for example, that induced
by nearby masses or tilt of the Raman lasers, are common for the
atoms in $m_F=+1$ and $m_F=-1$ and thus cancel in the final
comparison, and other disturbances, for example, that induced by the
AC-Stark shift or the Coriolis effect, can be neglected at the
present level of accuracy. The E$\rm{\ddot{o}}$tv$\rm{\ddot{o}}$s
ratio is finally given by
\begin{equation}
\eta_S  \equiv 2\frac{{{g_+} - {g_-}}}{{{g_+} + {g_-}}} =
2\frac{{\varphi _{{m_F} =  + 1}^d - \varphi _{{m_F} =
-1}^d}}{{\varphi _{{m_F} = + 1}^d + \varphi _{{m_F} =  - 1}^d}},
\label{Eq:3}
\end{equation}
where $\Delta \varphi _{{m_F}}^d$ is the corrected differential
results as listed in Table I. The resultant
E$\rm{\ddot{o}}$tv$\rm{\ddot{o}}$s ratio determined by this work is
$(-0.2\pm1.5)\times 10^{-5}$, which indicates that the violation of
WEP has not been observed at the level of $1.5\times 10^{-5}$ for
the atoms with different polarization orientations.

In conclusion, we have tested UFF with atoms in different spin
orientations based on a Mach-Zehnder-type atom interferometer, and
the violation of UFF isn't observed at the level of $1.5\times
10^{-5}$. This work represents the first atom interferometer which
simultaneously measures the gravity acceleration and magnetic-field
gradient, and also presents a direct test of spin-orientation
related spin-gravity couplings on quantum basis. The present
precision is limited by the fluctuation of the atomic fountain
arising from the atom launch procedure. The influence of this
fluctuation can be alleviated with a more homogeneous magnetic field
in future measurement, and in this situation the pulses separation
time can also be enlarged, which will effectively improve the
interferometer sensitivity. On the other hand, the standing optical
waves can be explored to manipulate the interfering of the atoms
(see Ref. \cite{27}, for example), in which case the internal state
of the atom doesn't change and thus the influence of the magnetic
field inhomogeneity is dramatically decreased. We anticipate a
better result for the UFF test with atoms in different spin
orientations using interferometers of this kind in future.

We thank Zhifang Xu for enlightening discussions. This work is
supported by the National Natural Science Foundation of China
(Grants No. 41127002, No. 11204094, and No. 11205064) and the
National Basic Research Program of China (Grant No. 2010CB832806).

\cleardoublepage

\end{document}